\documentclass{icrc2009}

\usepackage{graphicx}   
\usepackage{caption}    
\usepackage[font=footnotesize]{subfig} 
\usepackage{fixltx2e}
\usepackage{url}

\newcommand{\shorttitle}[1]%
{\markboth{Proceedings of the 31\MakeLowercase{$^{st}$} ICRC, {\L}\'{o}d\'{z} 2009}{#1} }
\newcommand{\etal}{\MakeLowercase{\textit{et al. }}} 

\begin{document}
\title{MAGIC observations of the distant quasar 3C279 during an optical outburst in 2007}

\author{\IEEEauthorblockN{Karsten Berger\IEEEauthorrefmark{1},
			  Pratik Majumdar\IEEEauthorrefmark{2},
                          Elina Lindfors\IEEEauthorrefmark{3},
                           Fabrizio Tavecchio\IEEEauthorrefmark{4} and
                           Masahiro Teshima\IEEEauthorrefmark{5}\\ 
                           on behalf of the MAGIC Collaboration\IEEEauthorrefmark{6}}\\
                          
\IEEEauthorblockA{\IEEEauthorrefmark{1}University of {\L }\'{o}d\'{z}, Department of Astrophysics, Pomorska 149/153, PL-90236 {\L }\'{o}d\'{z}, Poland}
\IEEEauthorblockA{\IEEEauthorrefmark{2}DESY Deutsches Elektronen-Synchrotron, D-15738 Zeuthen, Germany}
\IEEEauthorblockA{\IEEEauthorrefmark{3}Tuorla Observatory, University of Turku, FI-21500 Piikki\"o, Finland}
\IEEEauthorblockA{\IEEEauthorrefmark{4}INAF - National Institute for Astrophysics, I-00136 Rome, Italy}
\IEEEauthorblockA{\IEEEauthorrefmark{5}Max-Planck-Institut f\"ur Physik, D-80805 M\"unchen, Germany}
\IEEEauthorblockA{\IEEEauthorrefmark{6}the full author list can be found at:\\ http://wwwmagic.mppmu.mpg.de/collaboration/members/index.html}}

\shorttitle{K. Berger \etal MAGIC observations of 3C279 in 2007}
\maketitle

\begin{abstract}
 The flat-spectrum radio-quasar 3C279 (z=0.536) is the most distant object detected at very high energy (VHE) $\gamma$-rays. It is thus an important beacon for the study of the interaction of the VHE $\gamma$-rays with the Extra-galactic Background Light (EBL). Previous observations by EGRET showed a highly variable flux that can differ up to a factor of 100. In this paper results from an observation campaign with the MAGIC telescope during an optical flare in January 2007 will be presented and previous MAGIC results from 2006 will be summarized.
  \end{abstract}

\begin{IEEEkeywords}
 BL Lacertae objects: individual (3C279) - gamma-rays: observations; methods: data analysis
\end{IEEEkeywords}
 
\section{Introduction}

3C279 was the first blazar discovered as a $\gamma$-ray source with the Compton Gamma-Ray Observatory \cite{hartman1992} and the first --and up to now only-- flat-spectrum radio quasar (FSRQ) discovered to emit very high energy (VHE, defined here $>100$\,GeV) $\gamma$-rays \cite{science}.
With a redshift of 0.536 \cite{hewittburbridge1993} 3C279 is also the most distant of the VHE $\gamma$-ray emitting sources.

3C279 is one of the brightest sources in all wavelengths and its multiwavelength behavior and jet structure has been studied in great detail in several papers (e.g. \cite{hartman2001, chatterjee2008}). In this source the relativistic jet, which is the source of the radio to VHE $\gamma$-ray emission, is pointing very close to the line of sight (the angle varies, but is sometimes as small as $<0.5^\circ$ \cite{Jorstad2004}). The radio to optical emission is synchrotron radiation emitted by the relativistic electrons spiraling in the magnetic field of the jet. In this low energy regime the total flux density variations are well described by shocks propagating in the jet (e.g.
\cite{Lindfors2006}). The X-ray emission can be explained by the synchrotron self-Compton mechanism (SSC, e.g. \cite{hartman2001, Sikora2001}), where the synchrotron photons emitted by the jet act as seed photons for inverse Compton scattering. However, there is no consensus about the emission mechanism and site of the $\gamma$-ray and VHE $\gamma$-ray emission in 3C279. The emission can be, in principle, explained by both leptonic and hadronic models: the leptonic models mostly rely on external Compton (EC, e.g.
\cite{hartman2001, Albert}, invoking the inverse Compton scattering of external photons from broad-line region clouds, while in the hadronic models the VHE $\gamma$-ray photons are produced by proton synchrotron radiation (\cite{Muecke2001, Boettcher2009}).
The leptonic models are very sensitive to the site of the emission: the external Compton models relying on photons originating from broad line emission clouds are not efficient if the emitting blob is outside the broad line region and in the SSC models the $\gamma$-ray emission must originate from a different emission region than the main component of the synchrotron radiation in order to reproduce the observed $\gamma$-ray flux \cite{Boettcher2009}. It should also be noted, that independent of the emission mechanism the internal absorption cannot be neglected if the emission region is located inside the broad line region \cite{daniel}.

In this paper we present VHE $\gamma$-ray observations of 3C279 performed by the MAGIC-I telescope in January 2007. The observations were triggered by a large optical outburst of the source.
We present the analysis of the data, first results and their interpretation. For comparison and completeness we also summarize the 2006 results published in \cite{science}.

\section{Observations and Data Analysis}
 3C279 was observed with the MAGIC-I telescope \cite{Albert}, situated on the Canary Island of La Palma. With its 17\,m diameter tessellated reflector dish it is the largest imaging air Cherenkov telescope currently in operation. Its high quantum efficiency photomultiplier camera has a field of view of 3.5\,$^\circ$. MAGIC reaches a trigger threshold of 60\,GeV at zenith, an angular resolution of approximately 0.1\,$^\circ$ and an energy resolution above 150\,GeV of about 25$\%$. After the upgrade to an ultra-fast 2\,Ghz readout system in February 2007, the sensitivity could be considerably improved using the more precise shower timing information \cite{timing}, particularly close to the analysis threshold. Also data taken prior to this installation can take advantage of the timing information, albeit at a lower level (around 25$\%$) due to the lower timing resolution. 
 
  This paper focuses on data taken between January and April 2006 and in January 2007. Data taken during a multi-wavelength campaign in January 2009 will be discussed elsewhere. All data were taken in On/Off mode, where the telescope points directly at the source during the observations and Off data is collected later, at a point in the sky with very similar observational conditions. 3C279 was observed during ten nights in 2006, and during nine nights in 2007. 
  
  Simultaneous optical R-band observations were taken with the 1.03\,m telescope at the Tuorla Observatory (Finland) and the 35-cm KVA telescope on La Palma. The 2006 observations revealed that 3C 279 was in a high optical state during the MAGIC observations, a factor of 2 above its long-term baseline flux (host galaxy subtracted). At the beginning of 2007 the optical R-band flux reached a historically high flux, which triggered observations of the MAGIC-I telescope (Fig. 1). The observations started as soon as the moon light conditions allowed it (on January 14th 2007).

	 \begin{figure}[!t]
  \centering
  \includegraphics[width=2.5in]{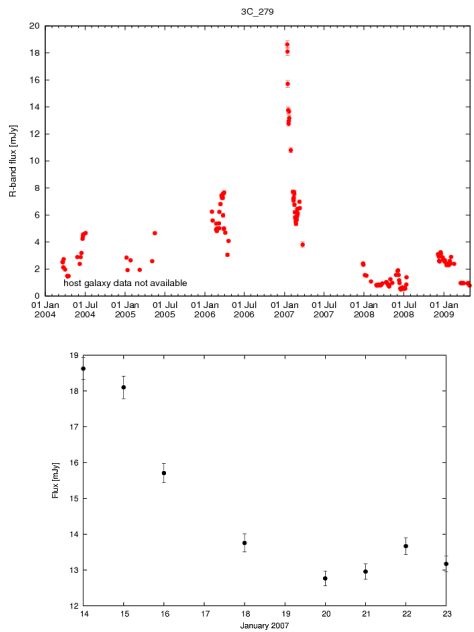}
  \caption{Top: Long term optical R-band light curve from the Tuorla monitoring program. The extraordinarly strong flaring emission in January 2007 is clearly visible. Bottom: Zoom into the optical light curve. Only nights with simultaneous MAGIC observations are shown.}
  \end{figure}

  The 2006 data were analysed using the standard MAGIC analysis and reconstruction software as described in \cite{Albert}. The 2007 data could, in addition, take advantage of the newly introduced timing analysis, described in \cite{timing}.

\section{Results}
	After background subtraction, a clear detection of 3C279 was found on the night of the 23rd of February 2006. The significance of the detection is 6.2$\sigma$ pre-trial, which is reduced to 5.8$\sigma$ when ten trials (one for each observation night) are considered. The 22nd of February shows a minor excess that is not significant enough to claim a detection. The complete light curve of the 2006 observation can be found in Fig. 2. A $\chi^2$ test determines the probability that the $\gamma$-ray flux was zero during all nights to be 2.3 x 10$^{-7}$, which corresponds to a significance of 5.0$\sigma$ in a Gaussian distribution. The observed differential VHE $\gamma$-ray spectrum (Fig. 3) can be reasonably described by a power law with a photon index of $\alpha=4.1\pm0.7_{stat}\pm0.2_{syst}$. The measured integrated flux above 100\,GeV on the 23rd of February is $(5.15\pm0.82_{stat}\pm1.50_{syst})$ x 10$^{-10}$cm$^{-2}$s$^{-1}$.
	
	\begin{figure}[!t]
  \centering
  \includegraphics[width=2.5in]{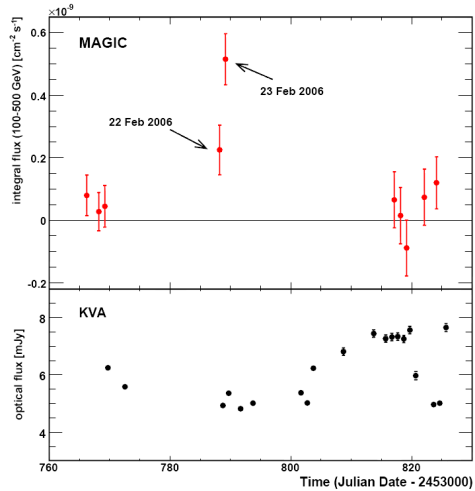}
  \caption{Light curves from the 2006 VHE $\gamma$-ray MAGIC (top) and optical R-band (bottom) observations (from \cite{science}). The optical flux of the host galaxy is negligeable. The long-term baseline of the optical flux is 3\,mJy.}
  \end{figure}

	 \begin{figure}[!t]
  \centering
  \includegraphics[width=2.5in]{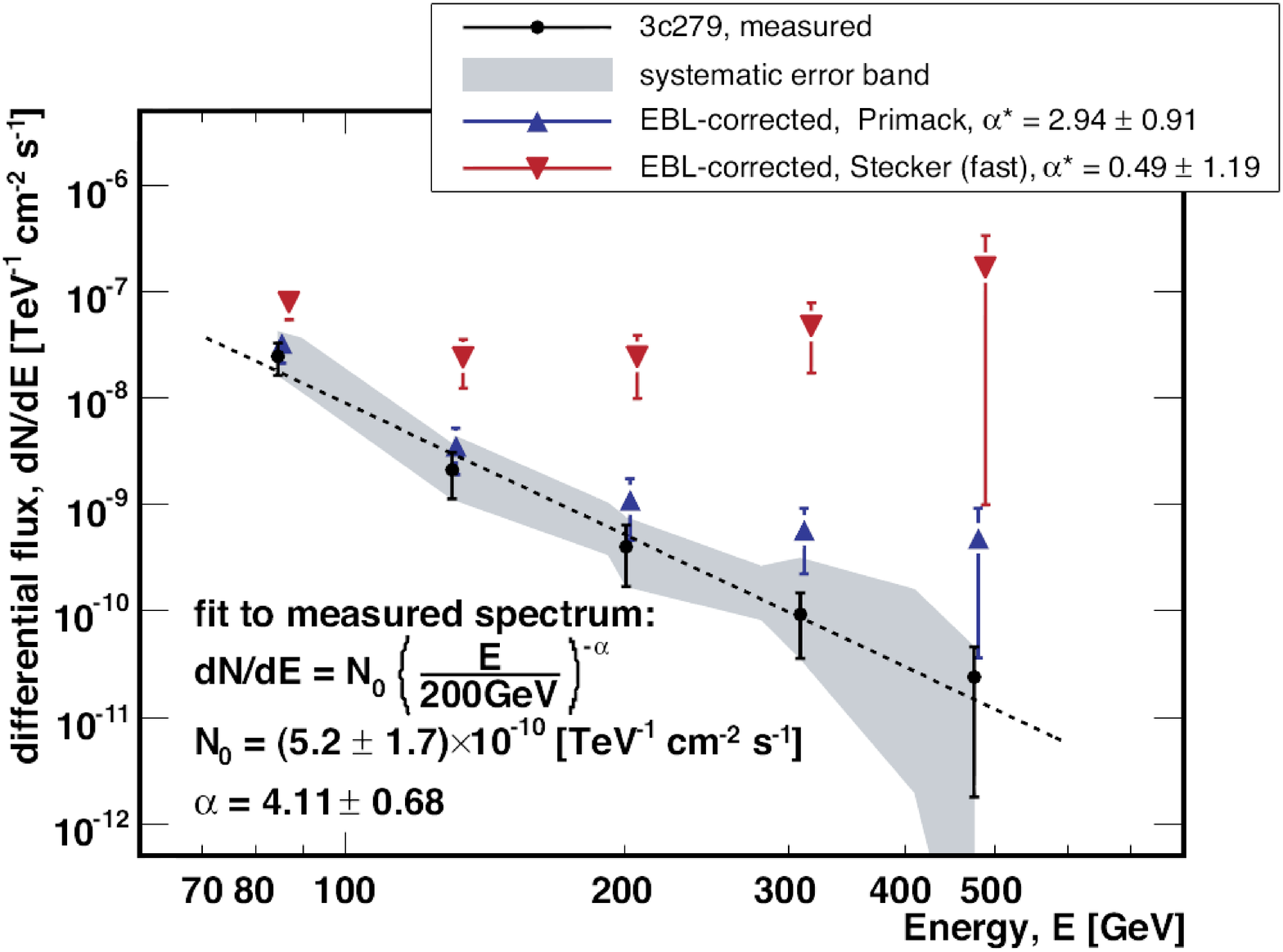}
  \caption{Spectrum of the VHE $\gamma$-ray flare that happened on the 23rd of February 2006. The black points correspond to the measured spectrum, while the red and blue triangles refer to the de-absorbed data points respectively, as described in the inlay. The grey shaded region represents the systematic error of the analysis and the dotted line refers to the fitted power law spectrum with a photon index $\alpha=4.1$.}
  \end{figure}

The analysis of the 2007 data set resulted in the discovery of a VHE $\gamma$-ray flare on the 16th of January (see Fig. 4). The significance of the excess corresponds to 5.6 $\sigma$ (5.2$\sigma$ after trials). The full analysis of the light curve is still underway. A preliminary analysis shows that the peak flux was reached on the 16th of January, several days after the peak of the optical flare. The spectral analysis of the data is ongoing and will be published at a later date.

	 \begin{figure}[!t]
  \centering
  \includegraphics[width=2.5in]{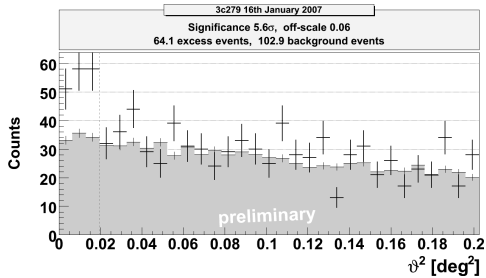}
  \caption{Reconstructed shower direction "$\vartheta^2$" for the Off (grey shaded area) and On (black crosses) data as observed on the 16th of January 2007 by MAGIC. The dotted line corresponds to the apriori defined signal region. An excess of 64 events is clearly visible in the On data when compared to the normalized Off data.}
  \end{figure}

\section{Discussion}

The discovery of VHE $\gamma$-ray emission from 3C279 has come as a great surprise, since its large distance results in strong absorption of the $\gamma$-rays due to the EBL. A confirmation of the previous detection was thus highly desirable. The second MAGIC observation campaign, triggered by an optical flare, achieved this confirmation, and thus firmly establishes 3C279 as a VHE $\gamma$-ray emitter. While the full analysis of the spectrum and light curve is still work in progress, preliminary results indicate a relatively short flare, which confirms that 3C279 is highly variable in the VHE $\gamma$-ray regime. A correlation study between the optical flux and the VHE $\gamma$-ray flux will follow as soon as the light curve has been finalized. The 2007 detection is one of several successful optical triggers that the MAGIC-I telescope has followed up (for more information see E. Lindfors \etal these proceedings).

 The detection of 3C279 at VHE $\gamma$-rays has interesting implications for the EBL. Since the EBL density is not very well known (due to uncertainties in the star formation rate and galaxy evolution) an independent measurement is highly desirable. We have used two extreme models to reconstruct the intrinsic source spectrum found in the 2006 data: a model by Primack \etal \cite{primack}, which is close to the lower limits set by galaxy counts \cite{galaxy1,galaxy2} and the "fast-evolution" model by Stecker \etal \cite{stecker}, which corresponds to the highest possible EBL. Both models are shown in Fig. 5 and will be referred to as "low" and "high", respectively. Using these models the intrinsic photon index can be calculted to be: $\alpha_{int}=2.9\pm0.9_{stat}\pm0.5_{syst}$ (for low absorption) and $\alpha_{int}=0.5\pm1.2_{stat}\pm0.5_{syst}$ (for high absorption). One can see that the $\it{high}$ absorption model results in an intrinsic photon index that is difficult to explain with an extrapolation of the EGRET data and with general constraints in the spectral energy distribution. However the $\it{low}$ absorption model gives a much more acceptable result.
 
 We assumed the hardest possible intrinsic photon index to be $\alpha_{int}=1.5$, which is the hardest value given for EGRET sources and additionally the hardest that can be obtained with classical leptonic emission mechanims. Using a model based on Kneiske \etal \cite{kneiske1,kneiske2} we can calculate an upper limit to the EBL density between 0.2 - 2$\mu$$m$ as shown in Fig. 5. These results confirm that the level of the EBL is close to the galaxy count limits, as was calculated using sources at lower redshift \cite{HESSebl}. It should be noted that alternative scenarios can produce intrinsic spectra with $\alpha_{int}<1.5$: SSC models with a narrow electron distribution \cite{narrowE} as well as internal absorption due to soft photons can harden the spectrum \cite{hardspectrum1, hardspectrum2, hardspectrum3, hardspectrum4}. The accurate modeling of the internal absorption in 3C279 performed in \cite{daniel}, however, concludes that no important hardening is expected within the MAGIC energy range.
 
 The emission process that is responsible for the observed VHE $\gamma$-ray spectrum remains uncertain. A spectrum corrected for $\it{low}$ level EBL absorption can be fitted by a one-zone SSC+EC model \cite{laura} (Fig. 6). Studies taking into account quasi-simultaneous optical and X-ray data \cite{boettcher1} conclude that neither one-zone SSC nor one-zone EC models can reproduce the observed SED \cite{boettcher2}.

	 \begin{figure}[!t]
  \centering
  \includegraphics[width=2.5in]{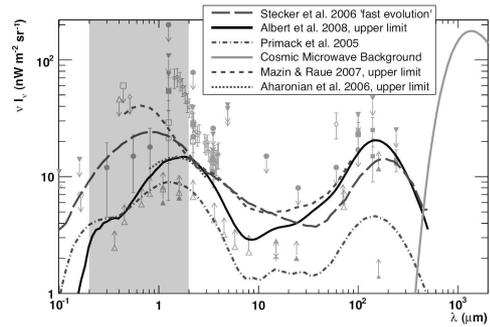}
  \caption{Measured EBL photon densitiy and model predictions for z=0 at various wavelengths. The black line corresponds to the upper limit derived from the detected VHE $\gamma$-ray spectrum of 3C279 in 2006 as discussed in \cite{science}. The area shaded in grey indicated the frequency range over which MAGIC is sensitive.}
  \end{figure}

	 \begin{figure}[!t]
  \centering
  \includegraphics[width=2.5in]{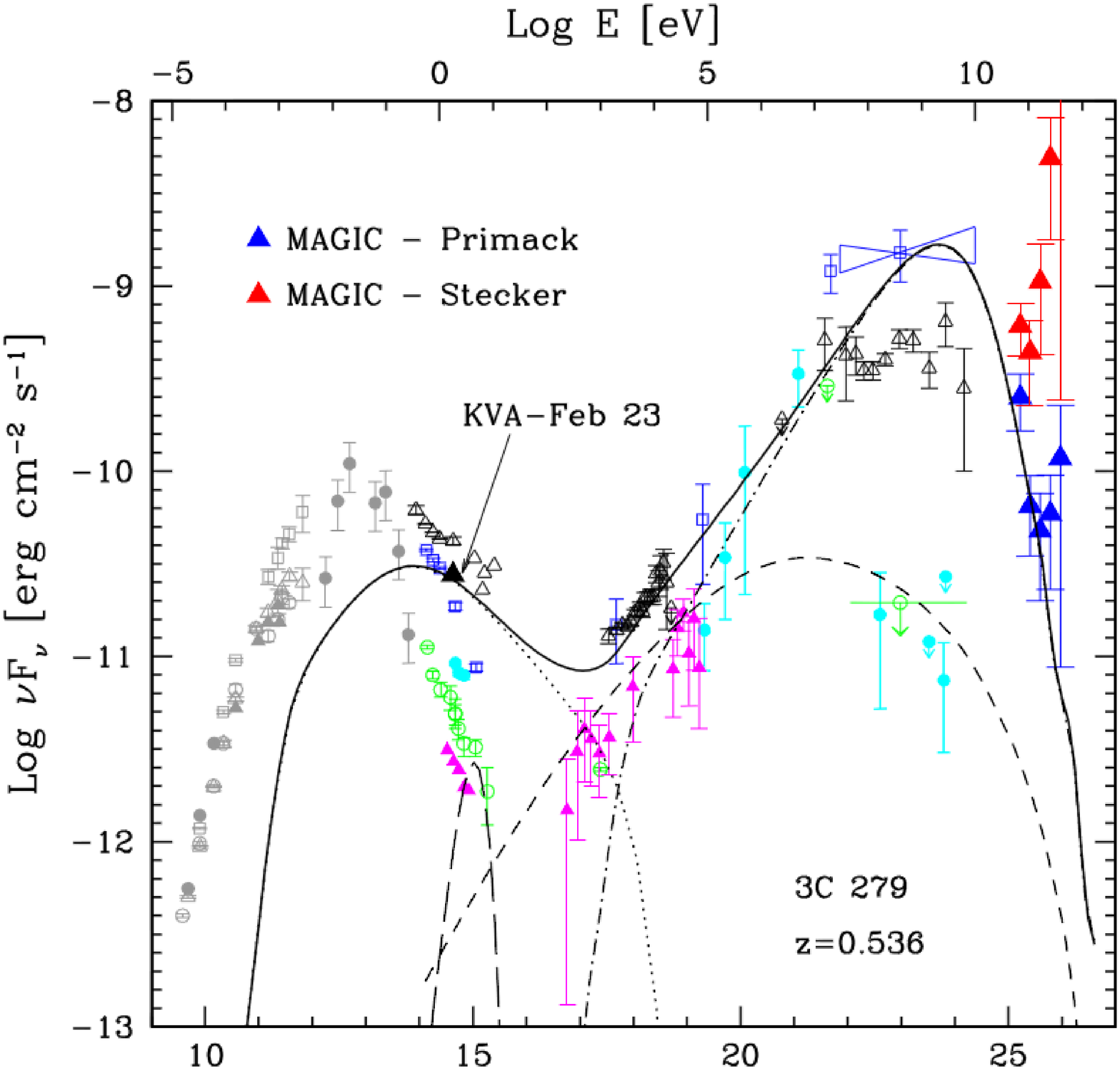}
  \caption{Spectral energy distribution of 3C279 for the years 1991 to 2003 (from \cite{3c279SED1} and \cite{3c279SED2}). The de-absorbed MAGIC data points from the 2006 observations are also shown. A $\it{low}$ and a $\it{high}$ EBL model has been used for the de-absorbtion, as discussed in the text. If no additional spectral component is present in the 2006 data, the $\it{low}$ EBL yields a much more plausible explanation of the data. An emission model has been fittet to the data (the black line illustrates the total model emission), which uses the $\it{low}$ EBL data points as well as the simultaneous optical KVA data point. The individual components of the model are also shown as dotted line (synchrotron radiation), long-dashed line (synchrotorn-self compton) and dot-dashed line (external Compton).}
  \end{figure}

\section*{Acknowledgments}
The MAGIC collaboration would like to thank the Instituto de Astrof\'{i}sica de Canarias for the excellent
working condition at the Observetorio del Roque de los Muchachos at La Palma. Major support from Germany's Bundesministerium f\"ur Bildung, Wissenschaft, Forschung und Technologie and Max-Planck-Gesellschaft, Italy's Istituto Nazionale di Fisica Nucleare (INFN) and Istituto Nazionale di Astrofisica (INAF), and Spain's Ministerio de Ciencia e Innovaci\'{o}n is gratefully acknowledged. The work was also supported by Switzerland's ETH Research grant TH34/043, Poland's Ministertwo Nauki i Szkolnictwa Wy\'{z}szego grant N N203 390834, and Germany's Young Investigator Program of the Helmholtz Gemeinschaft.

\end{document}